\documentclass[a4paper,11pt]{PoSLike}

\title{The strong-field QED experiment LUXE at the European XFEL and its detector challenges}
\ShortTitle{The strong-field QED experiment LUXE}

\author{Oleksandr Borysov \\
	Deutsches Elektronen-Synchrotron DESY,\\
	Notkestr. 85, 22607 Hamburg, Germany\\
    E-mail: \email{oleksandr.borysov@desy.de}}
\author{for the LUXE Collaboration}

\abstract{The LUXE experiment aims at studying high-field QED in electron-laser and photon-laser interactions, with the 16.5~GeV electron beam of the European XFEL and a laser beam with power of up to~350~TW. The experiment will measure the spectra of electrons and photons in non-linear Compton scattering where production rates in excess of $10^9$ are expected per~1~Hz bunch crossing. At the same time positrons from pair creation in either the two-step trident process or the Breit-Wheeler process will be measured, where the expected rates range from $10^{-3}$ to $10^3$ per bunch crossing, depending on the laser power and focus. These measurements have to be performed in the presence of low-energy high radiation-background. To meet these challenges, for high-rate electron and photon fluxes, the experiment will use Cherenkov radiation detectors, scintillator screens, sapphire sensors as well as lead-glass monitors for backscattering off the beam-dump. A four-layer silicon-pixel tracker and a compact electromagnetic tungsten calorimeter with GaAs sensors will be used to measure the positron spectra. The layout of the experiment and the expected performance under the harsh radiation conditions will be presented.}

\FullConference{%
  *** The European Physical Society Conference on High Energy Physics (EPS-HEP2021), ***\\
  *** 26-30 July 2021 ***\\
  *** Online conference, jointly organized by Universit\"{a}t Hamburg and the research center DESY ***
}

\usepackage{graphicx}
\usepackage{multirow}
\usepackage{textcomp}

\newcommand{\ecrit}{{\mathcal E}_\textrm{cr}}
\newcommand{\elaser}{{\mathcal E}_\textrm{L}}
\newcommand{\lambdabar}{{\mkern0.75mu\mathchar '26\mkern -9.75mu\lambda}}

\begin{document}

\section{Introduction}
Quantum electrodynamics (QED) with a strong electromagnetic field has been studied since the formulation of the QED. The scale of the strong field is known as a Schwinger critical field
\begin{equation}
\ecrit\equiv \frac{m_e^2 c^3}{e \hbar}\approx 1.3\times 10^{16} \ {\rm \frac{V}{cm}}\ .
\end{equation}
where~$m_e$ and~$e$ are electron mass and charge respectively, $c$~is a speed of light and~$\hbar$ is reduced Plank's constant. This field accelerates an electron to the energy equivalent to its mass at a distance of the Compton wavelength~$\lambdabar = \hbar/mc$, and leads to a possibility of spontaneous~e$^{+}$e$^{-}$~pair generation. A static filed of this magnitude is not reachable in the lab, but this regime of QED can be probed in collisions of high energy electrons or photons with an intense laser beam. Two strong-field phenomena have been extensively studied theoretically~\cite{boiling_point, BW_with_brems, pair_zero_mass}: non-linear Compton scattering and laser assisted~e$^{+}$e$^{-}$~pair production:
\begin{equation}
\label{eq_nonlin_compton}
e^{-} + n \gamma_L \rightarrow  e^{-} \gamma \, ,
\end{equation}
\begin{equation}
\label{eq_oppp}
\gamma + n \gamma_L \rightarrow  e^{+}  e^{-} \, ,
\end{equation}
where $n$ is the number of laser photons $\gamma_L$ participating in the process. A pioneer experiment to study these processes was performed at SLAC by E144 collaboration~\cite{e144_prd}.

\begin{figure}[h!]
	\begin{minipage}[t]{0.47\textwidth}
		\centering \includegraphics[width=0.99\columnwidth]{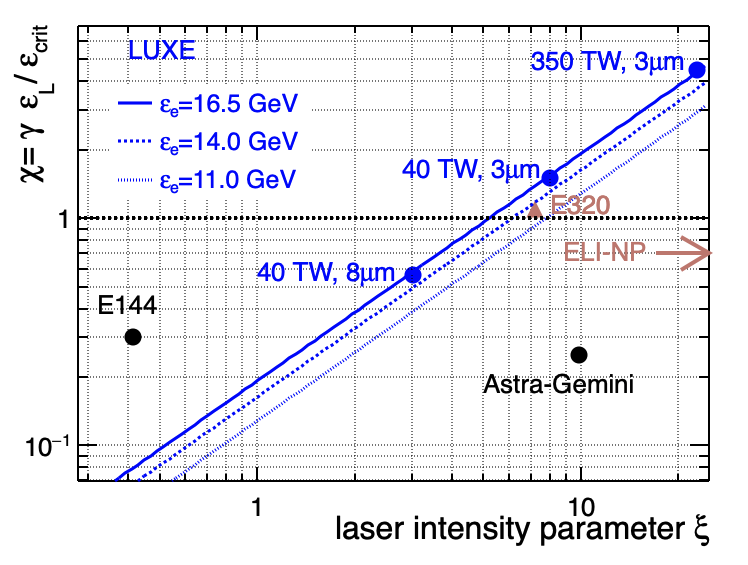}
		\caption{The $\chi$ and $\xi$ parameter space probed by various experiments. The three blue lines show the parameters accessible by LUXE for three possible electron beam energies. Two laser focus spot sizes are highlighted for the 40~TW (phase-0) laser and one for the 350~TW (phase-1) laser.}
		\label{fig_xi_chi_space}
	\end{minipage}\hfill
	\hspace{0.05\textwidth}
	\begin{minipage}[t]{0.47\textwidth}
		\centering \includegraphics[width=0.99\columnwidth]{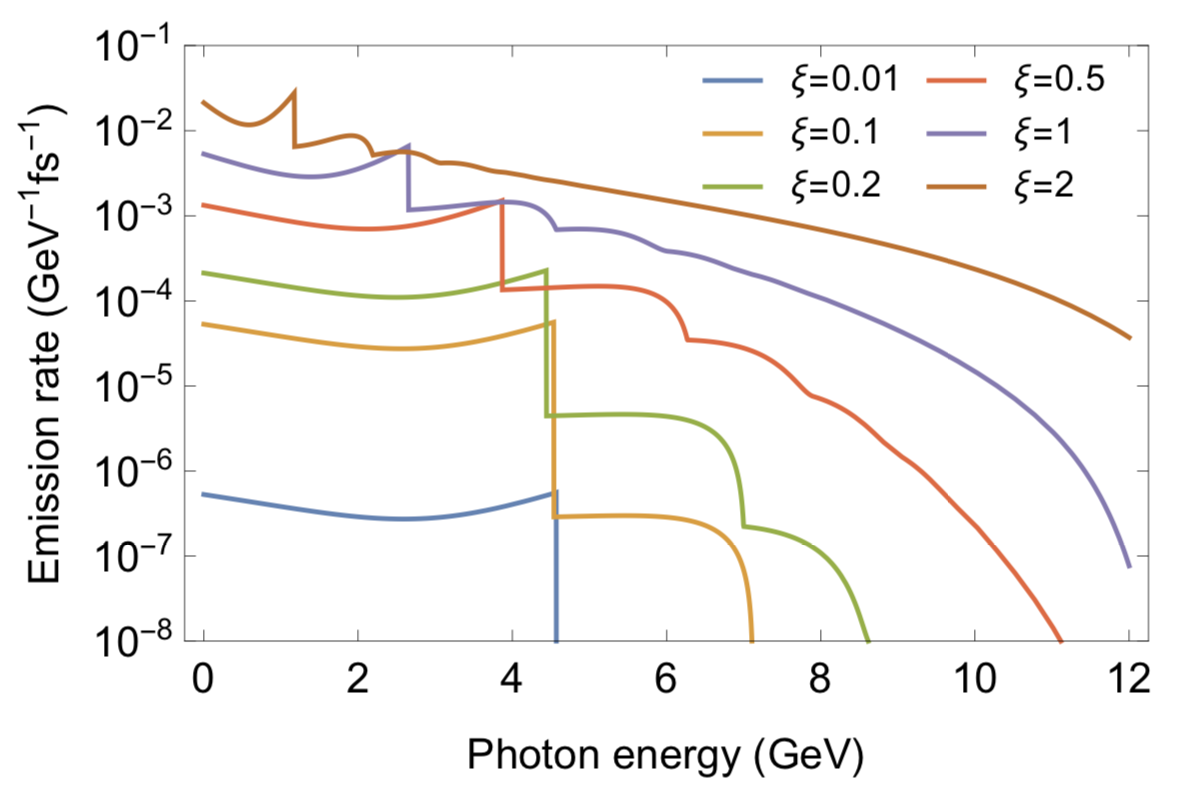}
		\caption{The rate of non-linear Compton scattering in collision of 16.5~GeV electrons with 800~nm laser and crossing angle 17.2\textdegree.}
		\label{fig_compton_cross_section}
	\end{minipage}\hfill
\end{figure}

The interaction between high energy particle and laser fields can be characterized by two dimensionless parameters. One of them is an intensity parameter related to the field strength:
\begin{equation}
\label{eq_xi}
\xi = \frac{e \elaser} {m_e c \omega_{L}}=\frac{m_e c^{2} \elaser}{\hbar \omega_L \ecrit}
\end{equation}
where $\elaser$ is the RMS electric field of the laser and $\omega_{L}$ its frequency. Another one is a quantum non-linearity parameter:
\begin{equation}
\label{eq_chii}
\chi = \frac{e \hbar}{m_{e}^3 c^5} \sqrt{(F_{\mu\nu} p^{\nu})^{2}} = \frac{{\mathcal E}^{\ast}}{\ecrit},
\end{equation}
where $F_{\mu\nu}$ is the field tensor, $ p^{\nu}$ is the four-momentum of the high energy particle and ${\mathcal E}^{\ast}$ is the electric field in particle's rest frame. The areas in this parameter space studied by E144, a recent experiment in Astra-Gemini laser facility~\cite{astra_gemini_2}, future planned experiments in European Extreme Light Infrastructure (ELI)~\cite{eli} and LUXE (Laser und XFEL~\cite{xfel_tdr} Experiment)~\cite{luxe_loi, luxe_cdr} at DESY are shown in figure~\ref{fig_xi_chi_space}. The possibility to investigate experimentally new domains in~$\xi$,~$\chi$ space is mainly determined by the tremendous progress in laser technologies achieved in recent decades. The proposed LUXE experiment aims to study non-perturbative QED processes (\ref{eq_nonlin_compton}),~(\ref{eq_oppp}) using the European XFEL electron beam and high power optical laser.

The differential cross section of non-linear Compton scattering~(\ref{eq_nonlin_compton}) as a function of the photon energy is presented in figure~\ref{fig_compton_cross_section} for different laser intensity parameters and the XFEL electron beam with an energy of 16.5~GeV. For the lowest shown value of~$\xi$ the spectrum is mainly determined by the electron interacting with a single photon of the laser field and exhibits a characteristic kinematic edge at~$E\approx4.6$~GeV. With increasing~$\xi$ from~0.01 to~2 the rate of Compton scattering increases by more than four orders of magnitude and includes the contributions from the processes where electron interacts with two, tree and more photons of the laser field. For~$\xi$ in the range~(0.01, 0.5) there are distinguishable first and second kinematic edges which correspond to one- and two-photon interactions. The values of the kinematic edges decrease as the laser intensity becomes higher. The measurements of the Compton scattering rate and spectra of electrons and photons with their specific feature is a part of the physics program of LUXE experiment. 
\begin{figure}[h!]
	\begin{minipage}[t]{0.47\textwidth}
		\centering \includegraphics[width=0.99\columnwidth]{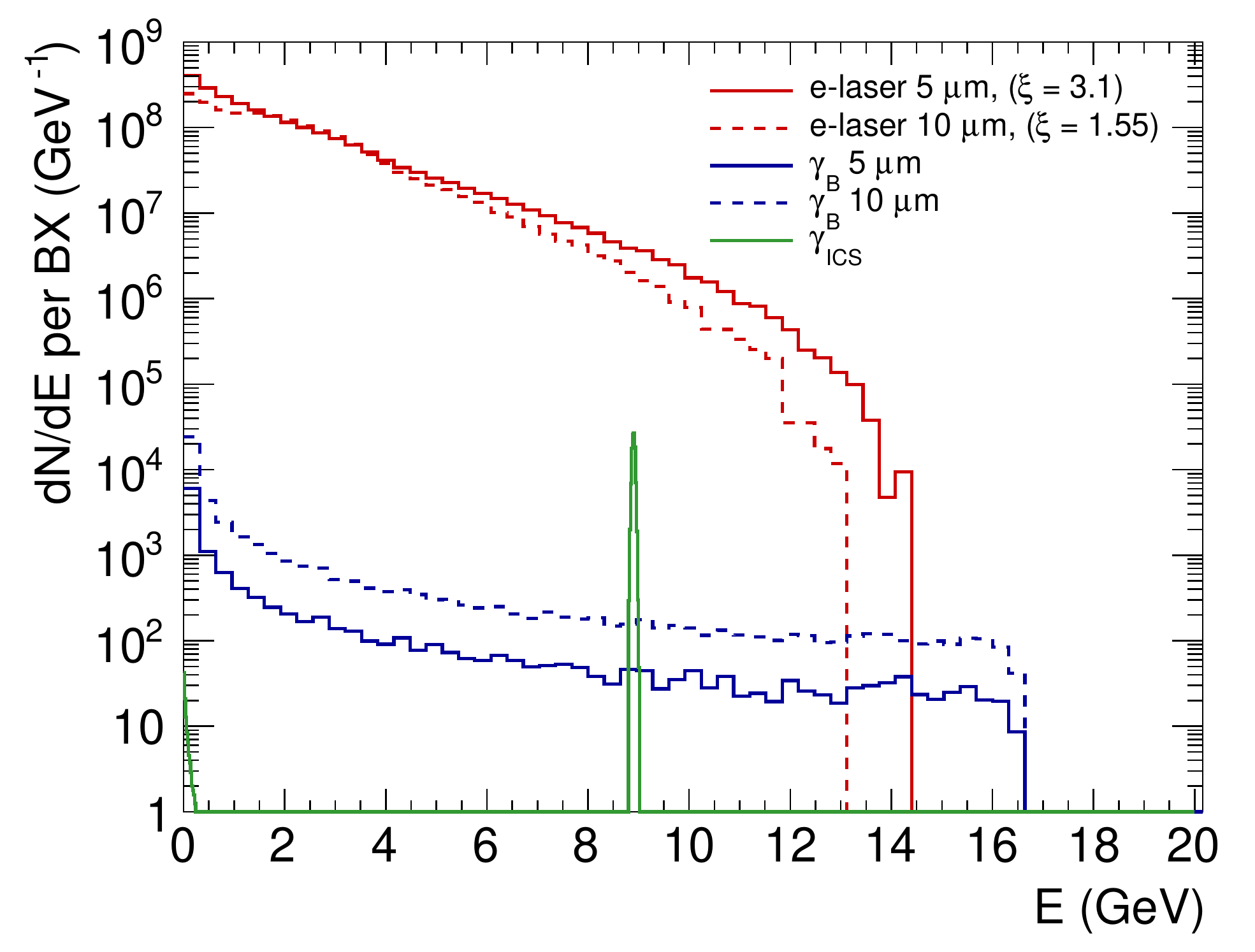}
		\caption{The spectra of photons at the interaction point obtained in MC simulations for different sources: non-linear Compton, bremsstrahlung and inverse Compton scattering (ICS).}
		\label{fig_gamma_source}
	\end{minipage}\hfill
	\hspace{0.05\textwidth}
	\begin{minipage}[t]{0.47\textwidth}
		\centering \includegraphics[width=0.99\columnwidth]{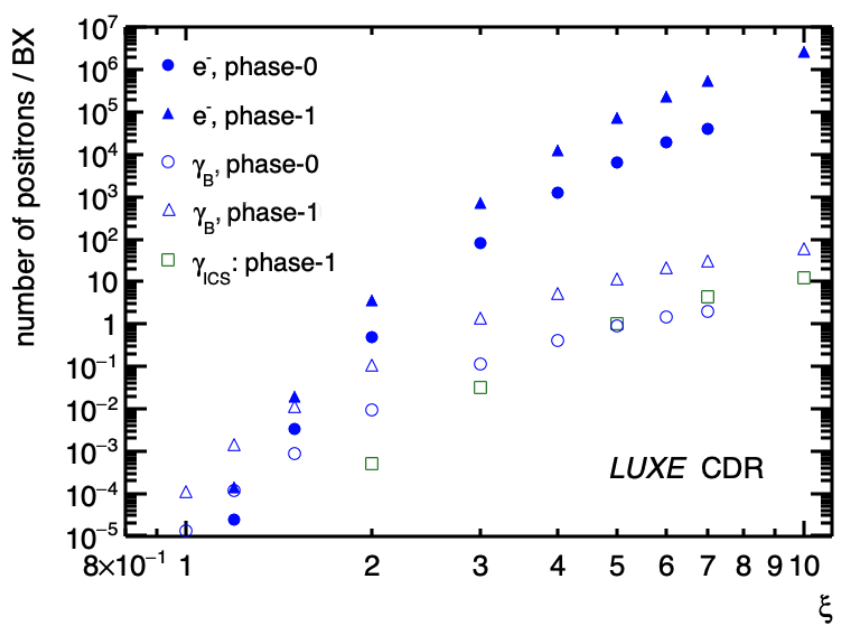}
		\caption{Number of positrons predicted by a MC simulation of non-linear Breit-Wheeler process in collisions of high energy photons with a laser pulse. Different markers correspond to different sources of the initial photons.}
		\label{fig_breit_wheeler}
	\end{minipage}\hfill
\end{figure}

Three possibilities are considered for studying the non-linear Breit-Wheeler process of electron positron pair production in interaction of high energy photon with the laser field~(\ref{eq_oppp}). The first possibility corresponds to the case, when the photon, produced in process~(\ref{eq_nonlin_compton}), propagating through the laser field, generates electron-positron pair in the same collision, where it was produced by electron beam. This intermediate photon in the process can be also virtual, then it is often referred as the "one-step" trident process. Other possibilities are to generate the high-energy photons significantly upstream of the main interaction point~(IP) via bremsstrahlung~\cite{boiling_point, BW_with_brems} and inverse Compton scattering~(ICS) with high frequency laser. The spectra of the photons for these three cases are shown in figure~(\ref{fig_gamma_source}). Bremsstrahlung photon spectra cover the range up to XFEL electron beam energy of 16.5~GeV, but the flux of the high energy part is substantially lower compared to the case of electron-laser collision. It is mainly limited by the geometrical constraints of the experiment~\cite{luxe_photon2019}. Using ICS, the pair production can be studied in collisions with an almost monochromatic photon beam, though its intensity and energy are relatively low, which results in comparatively low pair production rate.
\begin{table}[h!]
	\centering
	\begin{tabular}{|l|l|}
		\hline
		\textbf{Parameter} & \textbf{Value} \\
		\hline
		\hline
		Laser pulse energy (J)                       &  1.2 (phase-0), 10.0 (phase-1)   \\ 
		Laser transverse size, FWHM ($\mu$m)         &  8.0, 3.0    \\ 
		Laser pulse duration (fs)                    &  30     \\
		Laser repetition rate (Hz)                   &  1      \\
		Laser wavelength (nm)                        &  800  (1.5498~eV)   \\
		\hline
		Electron beam energy (GeV)                              &  16.5   \\ 
		Number of electrons  ($\times 10^{9}$)                  &  1.5   \\ 
		Electron beam transverse size, $\sigma_{x,y}$ ($\mu$m)  &  5 - 20   \\
		Electron beam duration (fs)                             &  80     \\
		Electron beam normalized emittance (mm~mrad)      &  1.4    \\
		\hline
		Crossing angle (rad)                                    & 0.35     \\ 
		\hline
	\end{tabular}
	\caption{LUXE laser and electron beam parameters at the IP.}
	\label{tab_luxe_beams}
\end{table} 
The numbers of electrons and positrons were estimated using strong field QED Monte Carlo~(MC) simulations~\cite{ptarmigan}, based on the locally monochromatic approximation (LMA)~\cite{qed_lma} and realistic electron and laser beam parameters as listed in table~(\ref{tab_luxe_beams}). The numbers are shown in figure~\ref{fig_breit_wheeler} as a function of the laser intensity parameter for different sources of high energy photons and their values span over ten orders of magnitude.
This report describes the results of simulation study and optimization of the LUXE setup for achieving its physics goals in challenging experimental environment. 
\section{LUXE setup}
\label{LUXE_setup}
As shown in final states in equations~\ref{eq_nonlin_compton} and~\ref{eq_oppp}, LUXE will measure electrons, positrons and photons and the setup of the experiment conceptually consists of electron and positron spectrometer, photon spectrometer and photon measuring subsystems. As the initial particles in these processes are different, the LUXE will run in two modes referred to hereafter as e-laser when initial particle is electron and $\gamma$-laser when it is high energy photon. 

\begin{figure}[h!]
	\centering
	\includegraphics[width=\textwidth]{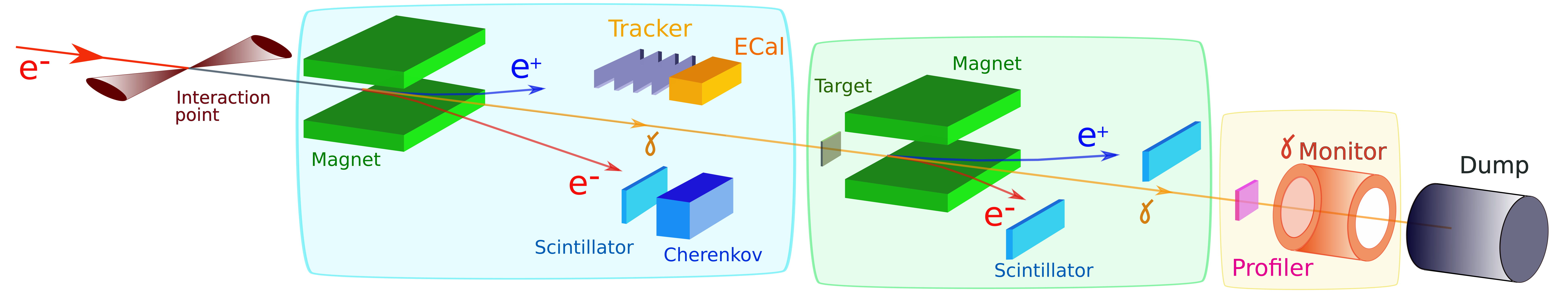}
	\caption{Diagram of the LUXE experiment layout. Shaded areas indicate electron positron spectrometer (blue), photon spectrometer (green) and photon detector (yellow) subsystems.}
	\label{fig_luxe_setup}
\end{figure}

A sketch of the LUXE experimental setup is presented in figure~\ref{fig_luxe_setup} for the e-laser mode. In this scenario the electron beam collides with the laser pulse at the IP located in the vacuum of the interaction chamber. Particles, produced in the collision, are confined to the narrow cone along the primary beam and propagate towards the electron and positron spectrometer which consists of a dipole magnet and a collection of detectors in electron and positron arms. The photon spectrometer, further downstream, consists of thin tungsten converter, dipole magnet and detectors for electron and positrons produced in the target. Most of the photons do not interact with the target material and are measured in the photon detector subsystem, where the gamma profiler determines the spatial distribution of the photons in the transverse plane and the gamma monitor is designed for measuring their numbers.

The optimization of the setup and detector performance study were carried out in simulation using the GEANT4~\cite{GEANT4_nima1} framework. The design of the detectors and chosen technologies are determined by the number of particles to be measured, the required accuracy and the background level. For the spectrometers this information is summarized in the table~\ref{tab_n_particles}.
\begin{table}[h!]
	\centering
	\begin{tabular}{||l|l||l|l||l||}
		\hline\hline
		\multicolumn{2}{||l||}{\textbf{Particle type}} & \multicolumn{2}{l||}{\textbf{Electron and positron spectrometer}} & \textbf{Photon spectrometer} \\
		\hline\hline
		\multirow{4}{*}{\textbf{e$^{-}$}} &            & \textbf{e$^{-}$~laser mode} & \textbf{$\gamma$~laser mode} & \\
		\cline{3-4}
		& Quantity   & 10$^{6}$~-~10$^{9}$ & 10$^{-2}$~-~10$^{1}$ & 10$^{5}$ \\
		\cline{2-5}
		& Detector   & scitillator screen     & pixel tracker followed  &  \multirow{2}{*}{scitillator screen} \\
		& technology & and Cherenkov detector & by calorimeter & \\
		\hline\hline
		\multirow{3}{*}{\textbf{e$^{+}$}} & Quantity   & 10$^{-2}$~-~10$^{4}$ & 10$^{-2}$~-~10$^{1}$ & 10$^{5}$ \\
		\cline{2-5}
		& Detector   & \multicolumn{2}{|l||}{\multirow{2}{*}{pixel tracker followed by calorimeter}} & \multirow{2}{*}{scitillator screen}\\
		& technology & \multicolumn{2}{|l||}{}                                                       & \\
		\hline\hline
	\end{tabular}
	\caption{Number of signal electrons and positrons per bunch collision in different detector subsystems of LUXE experiment and detector technologies chosen for their measurements.}
	\label{tab_n_particles}
\end{table}

The number of electrons per bunch crossing~(BX) varies substantially depending on the LUXE run mode and detector location in the experiment. There are several subsystems where the particle numbers exceed 10${^5}$. All of them are equipped with scintillator screens coupled with high resolution CMOS camera which takes a picture of the screen as it emits the light. This technology has been successfully used by AWAKE experiment~\cite{awake} and it provides position resolution better than~0.5~mm and up to~10~MGy radiation hardness for considered Tb-doped Gagolinium Oxysulfide screen. The sensitivity of the screen to the photons is essentially defined by their conversion, and since the thickness of the screen is about~0.5~mm, it is negligible which provides good rejection of the photon background. The electron arm of the first spectrometer also equipped with highly segmented gas Cherenkov detector. It provides complementary measurements of electron spectra with efficient rejection of the background represented by low energy charged particles.

\begin{figure}[h!]
	\begin{minipage}[t]{0.47\textwidth}
		\centering \includegraphics[width=0.99\columnwidth]{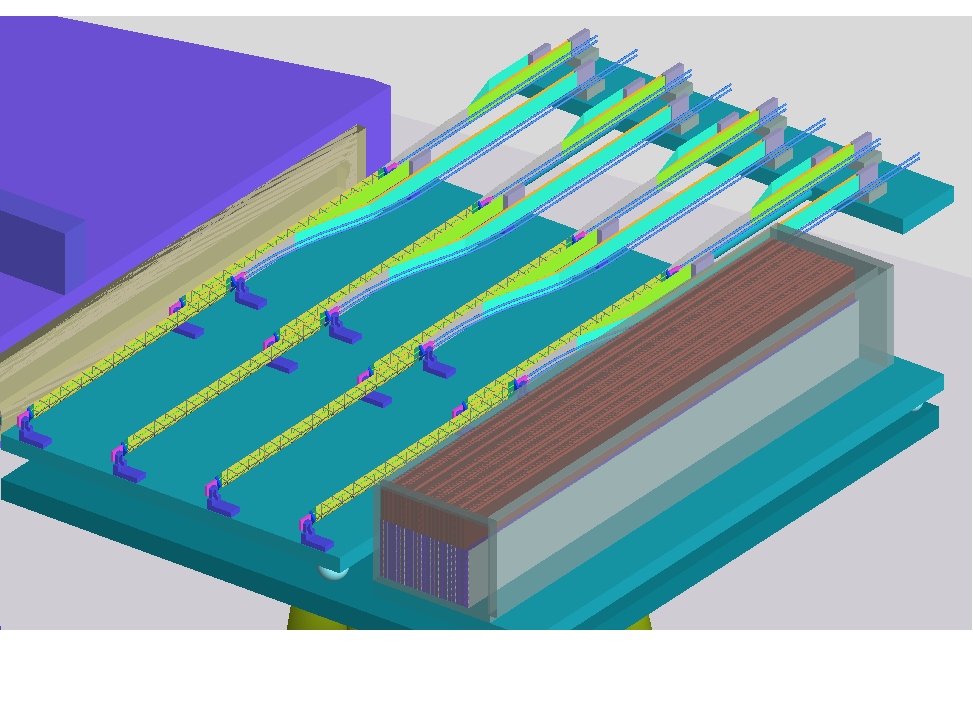}
        \caption{Positron detector GEANT4 model includes vacuum chamber window, four layers of tracking detectors followed by electromagnetic sampling calorimeter.} 
        \label{fig_tracker_g4}
	\end{minipage}\hfill
	\hspace{0.05\textwidth}
	\begin{minipage}[t]{0.47\textwidth}
		\centering \includegraphics[width=0.99\columnwidth]{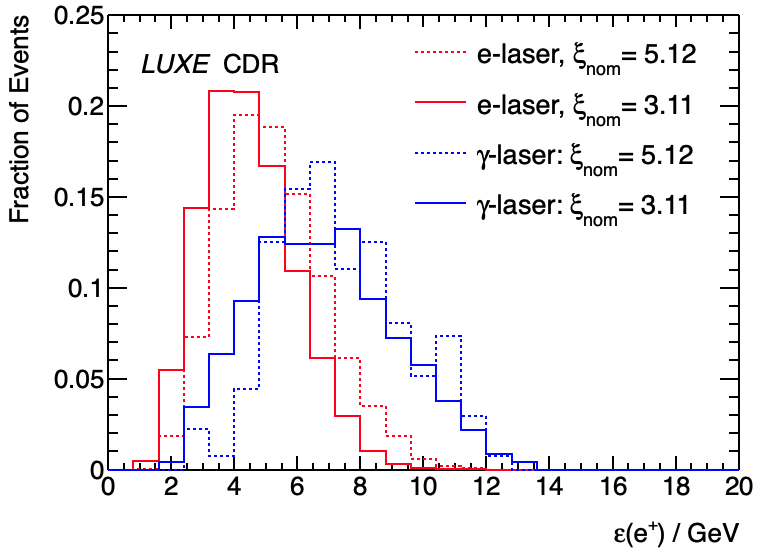}
        \caption{The positron energy spectrum for the e-laser and the $\gamma$-laser mode for phase-0.}
        \label{fig_positron_spectra}
	\end{minipage}\hfill
\end{figure}

The considered number of positrons per BX ranges from 10$^{-2}$ up to 10$^{4}$. The detection of small numbers of positrons requires excellent efficiency of the detector and robust background rejection. This background is substantial, considering the fact that the initial bunch comprises~1.5$\times$10$^{9}$ electrons and the beam is dumped in the experimental area. For this purpose the positron detector system consists of four layers of ALPIDE pixel sensors~\cite{alpide} followed by an ultracompact electromagnetic calorimeter~(fig.~\ref{fig_tracker_g4}). The pixel sensors are assembled in staves, two of which in each layer completely cover the expected spectra of positrons shown in figure~\ref{fig_positron_spectra}. The performance study of the tracker in GEANT4 simulation demonstrated efficiency close to~100\% and energy resolution better then~1\%. Ultracompact sampling electromagnetic calorimeter provides positron energy measurement and rejection of low energy charged particles background to which tracker is sensitive.

The gamma profiler consists of two 100~$\mu$m thick sapphire strip sensors and provides measurement of the photon distribution in the transverse plane with a resolution of about 5~$\mu$m. The sensors also have sufficient radiation hardness to operate during experiment run period. The gamma monitor is a calorimeter which measure the energy carried by particles backward from the photon beam dump. It was established in simulation studies that the energy deposited in calorimeter by back-scattered particles is nearly proportional to the number of photons in the beam which hit the dump. This allows us to measure the number of photons with 5\%~-~10\% accuracy.

For the $\gamma$-laser scenario, where photons are produced in bremsstrahlung, the initial electron beam hits the target upstream of the IP. The spectrometer located after the target, equipped with a scintillator screen and Cherenkov gas detectors, will register conversion electrons and positrons for estimating the number of produced photons. The number of electrons produced at the IP in the non-linear Breit-Wheeler process is identical to the number of positrons hence the same pixel tracking detector will be used for their registration. Four double layers of the tracker need to be installed on the movable platform capable of lifting the detector when the experiment switches to $\gamma$-laser mode.
\section{Summary}
LUXE~\cite{luxe_loi, luxe_cdr} at DESY proposes to extend the scientific scope of European XFEL to probe fundamental physics in the new regime of strong fields. Experimental study of laser assisted pair production and high intensity Compton scattering is feasible with European XFEL beam, combined with a multi-terawatt, high-intensity laser. Conceptual design study of LUXE experimental setup shows that the detector subsystems can be built using existing technologies for magnets, pixel tracking detectors, Cherenkov counters and calorimeters and that they can provide measurements with required accuracy while withstanding harsh experimental environment.
\section{Acknowledgements}
This work was in part funded by the Deutsche Forschungsgemeinschaft under Germany's Excellence Strategy -- EXC 2121 \textquotedblleft Quantum Universe\textquotedblright -- 390833306 and the German-Israel Foundation (GIF) under grant number 1492. It has benefited from computing services provided by the German National Analysis Facility (NAF).

\end{document}